\input epsf.tex


\font\titlefont = cmr10 scaled\magstep 4
 2
\font\sectionfont = cmr10
\font\littlefont = cmr5 
\font\eightrm = cmr8 

\def\ss{\scriptstyle} 
\def\sss{\scriptscriptstyle} 

\newcount\tcflag
\tcflag = 0  

\ifnum\tcflag = 0 \magnification = 1200 \fi  

\global\baselineskip = 1.2\baselineskip 
\global\parskip = 4pt plus 0.3pt 
\global\abovedisplayskip = 18pt plus3pt minus9pt
\global\belowdisplayskip = 18pt plus3pt minus9pt
\global\abovedisplayshortskip = 6pt plus3pt
\global\belowdisplayshortskip = 6pt plus3pt

\def\barsoff{\overfullrule=0pt}


\def\endignore{}
\def\ignore #1\endignore{} 

\newcount\dflag
\dflag = 0


\def\monthname{\ifcase\month 
\or January \or February \or March \or April \or May \or June%
\or July \or August \or September \or October \or November %
\or December 
\fi}

\newcount\dummy
\newcount\minute  
\newcount\hour
\newcount\localtime
\newcount\localday
\localtime = \time
\localday = \day

\def\advanceclock#1#2{ 
\dummy = #1
\multiply\dummy by 60
\advance\dummy by #2
\advance\localtime by \dummy
\ifnum\localtime > 1440 
\advance\localtime by -1440
\advance\localday by 1
\fi}

\def\settime{{\dummy = \localtime %
\divide\dummy by 60%
\hour = \dummy 
\minute = \localtime%
\multiply\dummy by 60%
\advance\minute by -\dummy 
\ifnum\minute < 10 
\xdef\spacer{0} 
\else \xdef\spacer{} 
\fi %
\ifnum\hour < 12 
\xdef\ampm{a.m.} 
\else 
\xdef\ampm{p.m.} 
\advance\hour by -12 %
\fi %
\ifnum\hour = 0 \hour = 12 \fi 
\xdef\timestring{\number\hour : \spacer \number\minute%
\thinspace \ampm}}}



\def\endtitle{}
\def\title#1\endtitle{\vskip.5in\titlefont
\global\baselineskip = 2\baselineskip 
#1\vskip.4in
\baselineskip = 0.5\baselineskip\rm}
 
\def\endauthors{}
\def\authors#1\endauthors{#1}

\def\endabstract{}
\def\abstract#1\endabstract{\vskip .3in%
\centerline{\sectionfont\bf Abstract}%
\vskip .1in
\noindent#1}

\def\nopageonenumber{\footline={\ifnum\pageno<2\hfil\else
\hss\tenrm\folio\hss\fi}}  

\newcount\nsection 
\newcount\nsubsection 

\def\section#1{\global\advance\nsection by 1
\nsubsection=0
\bigskip\noindent\centerline{\sectionfont \bf \number\nsection.\ #1}
\bigskip\rm\nobreak}

\def\subsection#1{\global\advance\nsubsection by 1
\bigskip\noindent\sectionfont \sl \number\nsection.\number\nsubsection)\
#1\bigskip\rm\nobreak}

\def\topic #1{{\medskip\noindent $\bullet$ \it #1:}} 
\def\endtopic{\medskip}

\def\appendix#1#2{\bigskip\noindent%
\centerline{\sectionfont \bf Appendix #1.\ #2} 
\bigskip\rm\nobreak} 


\newcount\nref 
\global\nref = 1 

\def\therefs{} 


\def\ref#1#2{\xdef #1{[\number\nref]} 
\ifnum\nref = 1\global\xdef\therefs{\item{[\number\nref]} #2\ } 
\else
\global\xdef\oldrefs{\therefs}
\global\xdef\therefs{\oldrefs\vskip.1in\item{[\number\nref]} #2\ }%
\fi%
\global\advance\nref by 1
}

\def\listrefs{\vfill\eject\section{References}\therefs}


\newcount\nfoot 
\global\nfoot = 1 

\def\foot#1#2{\xdef #1{(\number\nfoot)} 
\hskip -0.2cm ${}^{\number\nfoot}$ 
\footnote{}{\vbox{\baselineskip=10pt
\eightrm \hskip -1cm ${}^{\number\nfoot}$ #2}}
\global\advance\nfoot by 1
}


\newcount\nfig 
\global\nfig = 1
\def\thefigs{} 

\def\figure#1#2{\xdef #1{(\number\nfig)}
\ifnum\nfig = 1\global\xdef\thefigs{\item{(\number\nfig)} #2\ }
\else
\global\xdef\oldfigs{\thefigs}
\global\xdef\thefigs{\oldfigs\vskip.1in\item{(\number\nfig)} #2\ }%
\fi%
\global\advance\nfig by 1 } 

\def\fig#1{\xdef #1{(\number\nfig)}
\global\advance\nfig by 1 } 


\newcount\ntab
\global\ntab = 1

\def\table#1{\xdef #1{\number\ntab}
\global\advance\ntab by 1 } 


\newcount\cflag
\newcount\nequation
\global\nequation = 1
\def\eqlabel{(1)}

\def\nexteqno{\ifnum\cflag = 0
\global\advance\nequation by 1
\fi
\global\cflag = 0
\xdef\eqlabel{(\number\nequation)}}

\def\lasteqno{\global\advance\nequation by -1
\xdef\eqlabel{(\number\nequation)}}

\def\label#1{\xdef #1{(\number\nequation)}
\ifnum\dflag = 1
{\escapechar = -1
\xdef\draftname{\littlefont\string#1}}
\fi}

\def\clabel#1#2{\xdef\eqlabel{(\number\nequation #2)}
\global\cflag = 1
\xdef #1{\eqlabel}
\ifnum\dflag = 1
{\escapechar = -1
\xdef\draftname{\string#1}}
\fi}

\def\cclabel#1#2{\xdef\eqlabel{#2)}
\global\cflag = 1
\xdef #1{\eqlabel}
\ifnum\dflag = 1
{\escapechar = -1
\xdef\draftname{\string#1}}
\fi}


\def\eeq{}

\def\eqnn #1\eeq{$$ #1 $$}

\def\eq #1\eeq{
\ifnum\dflag = 0
{\xdef\draftname{\ }}
\fi 
$$ #1
\eqno{\eqlabel \rlap{\ \draftname}} $$
\nexteqno}







\def\eqa #1\eeq{
\ifnum\dflag = 0
{\xdef\draftname{\ }}
\fi 
$$ \eqalignno{ #1 } $$
\global\cflag = 0}


\def\etc{{\it etc.\/}}

\def\vs{{\it vs.\/}}


\def\anp#1#2#3{{\it Ann.\ Phys. (NY)} {\bf #1} (19#2) #3}

 \def\app#1#2#3{{\it Act.\ Phys. \ Polon.} {\bf B#1}, (19#2) #3}

\def\cmp#1#2#3{{\it Comm.\ Math.\ Phys.} {\bf #1} (19#2) #3}

\def\mpla#1#2#3{{\it Mod.\ Phys.\ Lett.} {\bf A#1}, (19#2) #3}

\def\npb#1#2#3{{\it Nucl.\ Phys.} {\bf B#1} (19#2) #3}
\def\plb#1#2#3{{\it Phys.\ Lett.} {\bf #1B} (19#2) #3}

\def\prd#1#2#3{{\it Phys.\ Rev.} {\bf D#1} (19#2) #3}


\global\nulldelimiterspace = 0pt



\def\frac#1#2{{{#1} \over {#2}}\,}  
\def\hf{{1\over 2}}



\def\Square{{\vbox {\hrule height 0.6pt\hbox{\vrule width 0.6pt\hskip 3pt
        \vbox{\vskip 6pt}\hskip 3pt \vrule width 0.6pt}\hrule height 0.6pt}}}
\def\Asl{\hbox{/\kern-.7500em\it A}} 
\def\Dsl{\hbox{/\kern-.6700em\it D}} 
\def\dsl{\hbox{/\kern-.5300em$\partial$}}
\def\pxpsl{\hbox{/\kern-.5600em$p$}}
\def\sslsh{\hbox{/\kern-.5300em$s$}}
\def\epssl{\hbox{/\kern-.5100em$\epsilon$}}
\def\delsl{\hbox{/\kern-.6300em$\nabla$}}
\def\lxpsl{\hbox{/\kern-.4300em$l$}}
\def\elxpsl{\hbox{/\kern-.4500em$\ell$}}
\def\kxpsl{\hbox{/\kern-.5100em$k$}}
\def\qxpsl{\hbox{/\kern-.5000em$q$}}
\def\sla#1{\raise.15ex\hbox{$/$}\kern-.57em #1}



\def\roughly#1{\mathrel{\raise.3ex\hbox{$#1$\kern-.75em\lower1ex\hbox{$\sim$}}}}





\def\Scl{{\cal L}}


\def\ssf{{\sss F}}


\def\pmb#1{\setbox0=\hbox{#1}%
\kern-.025em\copy0\kern-\wd0
\kern.05em\copy0\kern-\wd0
\kern-.025em\raise.0433em\box0}   


\font\jlgtenbrm=cmbx10
\font\jlgtenbit=cmmib10
\font\jlgtenbsy=cmbsy10
\font\jlgsevenbrm=cmbx10 at 7pt
\font\jlgsevenbsy=cmbsy10 at 7pt
\font\jlgsevenbit=cmmib10 at 7pt
\font\jlgfivebrm=cmbx10 at 5pt
\font\jlgfivebsy=cmbsy10 at 5pt
\font\jlgfivebit=cmmib10 at 5pt
\newfam\jlgbrm

\textfont\jlgbrm=\jlgtenbrm
\scriptfont\jlgbrm=\jlgsevenbrm
\scriptscriptfont\jlgbrm=\jlgfivebrm
\newfam\jlgbit

\textfont\jlgbit=\jlgtenbit
\scriptfont\jlgbit=\jlgsevenbit
\scriptscriptfont\jlgbit=\jlgfivebit
\newfam\jlgbsy

\textfont\jlgbsy=\jlgtenbsy
\scriptfont\jlgbsy=\jlgsevenbsy
\scriptscriptfont\jlgbsy=\jlgfivebsy
\newcount\jlgcode
\newcount\jlgfam
\newcount\jlgchar
\newcount\jlgtmp
\def\bolded#1{
        \jlgcode\the#1 \divide\jlgcode by 4096
        \jlgtmp\the\jlgcode \multiply\jlgtmp by 4096
        \jlgfam\the#1 \advance\jlgfam by -\the\jlgtmp
        \divide\jlgfam by 256
        \jlgtmp\the\jlgcode \multiply\jlgtmp by 16
	\advance\jlgtmp by \the\jlgfam
	\multiply\jlgtmp by 256
        \jlgchar\the#1 \advance\jlgchar by -\the\jlgtmp
        \advance\jlgfam by \the\jlgbrm
        \jlgtmp\the\jlgcode
        \multiply\jlgtmp by 16
        \advance\jlgtmp by \the\jlgfam
        \multiply\jlgtmp by 256
        \advance\jlgtmp by \the\jlgchar
        \mathchar\the\jlgtmp
}


\def\Tr{\mathop{\rm Tr}}
\def\det{\mathop{\rm det}}



\def\Avg#1{\left\langle #1 \right\rangle}





\nopageonenumber
\baselineskip = 18pt
\barsoff



\def\mn{{\mu\nu}}

\def\eff{{\rm eff}}

\def\L{\Lambda}


\line{hep-th/9808107  \hfil McGill-98/17. }

\vskip .7in
\title
\centerline{Massive-Scalar Effective Actions}
\centerline{on Anti-de Sitter Spacetime}
\endtitle

\vskip 0.2in
\authors
\centerline{M. Kamela and C.P. Burgess}
\vskip .1in
\centerline{\it Physics Department, McGill University}
\centerline{\it 3600 University St., Montr\'eal, Qu\'ebec, Canada,
H3A 2T8.}
\endauthors

\abstract
\vbox{\baselineskip 15pt
Closed forms are derived for the effective actions for free,
massive spinless fields in anti-de Sitter spacetimes in arbitrary
dimensions. The results have simple expressions in terms of
elementary functions (for odd dimensions) or
multiple Gamma functions (for even dimensions).
We use these to argue against the quantum
validity of a recently-proposed
duality relating such theories with differing masses and
cosmological constants. }
\endabstract


\vfill\eject

\section{Introduction}

\ref\maldacena{J. Maldacena, hep-th/9712200}

In this note we give explicit expressions for the
effective actions for free, massive scalar fields propagating
within anti-de Sitter (AdS) spacetimes of arbitrary
dimension. Besides their intrinsic interest as exact expressions
for quantum systems interacting with nontrivial gravitational
fields, or as the first terms in a derivative expansion for
more complicated backgrounds,
these actions may also have applications to the calculation
of quantum effects within cosmologically-interesting spacetimes.
Remarkably, their supersymmetric extensions in five-dimensions may
prove useful for study of large-$N$ corrections to nonabelian
gauge theories, in view of the recently-proposed duality between
these theories and AdS supergravity in
five dimensions \maldacena.

\ref\avisetal{S. Avis,  C. Isham, D. Storey, \prd{18}{78}{3565}.}

\ref\breitfreed{P. Breitenlohner, D. Freedman, \anp{144}{82}{249} }

\ref\coleman{S. Coleman,  \npb{259}{85}{170}   }

\ref\duffandco{S. Christensen, M. Duff, \npb{170}{80}{480}  }

\ref\candelasraine{P. Candelas, D. Raine, \prd{12}{75}{965}  ,
\prd{15}{76}{1494} }

\ref\allenjacobson{B. Allen, T. Jacobson, \cmp{103}{86}{669}   }

\ref\burgesslutken{C. Burgess, A. Lutken, \plb{153}{84}{137}   }

\ref\camp{R. Camporesi, \prd{43}{91}{3958} }

\ref\odint{S. Buchbinder, S. Odintsov,  \app{18}{87}{237}  }

\ref\allen{B. Allen, \npb{226}{83}{228}   }

\ref\camphig{R. Camporesi, A. Higuchi, \prd{47}{93}{3339} }

Our calculations extend a number of similar calculations
which have been performed by others in the past. Much of the
early interest was motivated by the questions
of principle which arise when quantizing fields in these
spacetimes \avisetal, \breitfreed, and by vacuum-stability
\breitfreed, \coleman\ and divergence \duffandco\
issues associated with the appearance of AdS spacetimes
as supersymmetric vacua in extended-supergravity models.
Starting very early, the maximal symmetry of these spacetimes was
harnessed to perform explicit effective-action calculations
for scalar fields in both de Sitter \candelasraine,
\allenjacobson, and anti-de Sitter
\allenjacobson, \burgesslutken,
\camp, \odint, as
well as calculations of the functional determinants
which arise in higher-spin calculations \allen, \camphig.
The main advantage of our expressions over those in the literature
is their validity for general spacetime dimension. For odd
dimensions the results may be expressed in closed form
using elementary functions. For even dimensions we also
obtain closed-form results in terms of a
class of special functions --- the multiple gamma functions,
$\{G_n\}$ --- whose properties have been extensively studied.

\ref\claim{J. Cruz,  {\it Hidden conformal symmetry of a massive scalar
field in
$AdS_2$}, hep-th/9806145.}

Although we had performed the calculations we describe here for other
applications in mind, one of our motivations for
reporting the results now is the recent claim \claim\
for the existence of a duality relating
scalar field theories of mass $m^2 = 0$ and $m^2 = R$
in two-dimensional anti-de Sitter spacetimes with Ricci scalar $R$.
We believe our calculations provide evidence against this
duality existing as a quantum symmetry.

Our presentation is organized in the following way.
In \S1\ we briefly review some properties of anti-de Sitter
spaces which are useful for obtaining the effective action.
\S2\ contains our main result: the derivation of the
scalar-field effective action in an AdS space-time of arbitrary
dimension, $n$. \S3\ specializes this result to various
cases of particular interest. For even dimensions we
display results for $n=2$ and
$n=4$, where we reproduce previous calculations. (For
$n=4$ we also give, in passing, an expression of the results
for general spin in terms of the multiple Gamma functions.)
We also present the odd-dimensional cases $n=3,5$ and 7,
which have not been previously calculated. Since our
results are valid for arbitrary scalar masses and cosmological
constants, they bear on the issue of the existence of
duality transformations relating different values of these
parameters. The duality analysis is the topic of \S4.
Finally, we gather
some useful definitions and properties of the multiple Gamma
functions in an appendix.

\section{Scalar Fields on Anti-de Sitter Spacetime}

\ref\mtw{See, for instance: C. Misner, K. Thorne, J. Wheeler,
{\it Gravitation}, Freeman, 1970.}

\ref\weinberg{S. Weinberg, {\it Gravitation and
Cosmology} , Chapter 13, John Wiley \& Sons, New York, 1972.}

An $n$-dimensional spacetime which admits $\frac12 \,n(n+1)$
Killing vectors is said to be maximally
symmetric \mtw, \weinberg. The Riemann
curvature tensor for any such spacetime may be written in the
following way:\foot\conventions{Our conventions are those
of ref.~\weinberg.}
\eq
R_{\lambda \rho \sigma \nu} = K (g_{\sigma \rho } g_{\lambda \nu}
- g_{\nu \rho} g_{\lambda \sigma}) \quad \quad R = -\,  n(n-1) K ,
\label\curvature
\eeq
where $K$ a real constant.
The possible maximally-symmetric spaces
which can be entertained may be characterized by the signatures
of their metrics as well as the sign of their Ricci scalar $R$
(or, $K$). In our conventions anti-de Sitter space
is the pseudo-Riemanninan space for which $R > 0$, and
so for which $K = - \lambda^2 < 0$.

Quantization of scalar field theory on AdS
spacetime involves additional complications over those
which arise for flat Minkowski space. Besides
unrolling the compact time direction and working on the
Universal Covering Space, the tricky feature about
field quantization on AdS is connected with this
spacetime not being globally
hyperbolic \avisetal, \breitfreed.
(That is, in order for the scalar-field equations
to formulate a well-posed boundary-value problem,
boundary information is required
on a time-like surface at spatial infinity in addition
to the usual initial conditions which would have
been sufficient in Minkowski space.) This
complications lead to the existence of more than
one Fock vacuum for the quantum field theory. As a consequence,
different physical situations can lead to different boundary
conditions, and so to different quantum field theories.

Given a scalar quantum field on AdS spacetime, our
goal is to compute the scalar-field contribution, $\Sigma$,
to the effective action.
This is given by the following path integral:
\eq
\eqalign{
e^{i\Sigma(g_\mn,m^2)} &= \int [D X]_{g_{\mu \nu}}
\; \exp \left[-\;  \frac{i}{2} \int d^n x
\sqrt{-g} \; X \left(  -\Square_g + m^2 \right) X \right] \cr
&= \left[{\det}^\prime \left( - \Square_g + m^2 \right)\right]^{-1/2} , \cr}
\label\scalarPI
\eeq
where $\Square_g := \frac{1}{\sqrt{-g}} \partial_\mu 
\left( g^{\mu \nu} \sqrt{-g} \partial_\nu \right)$ 
is the usual Laplacian operator acting on scalar fields, and
the prime in the second equality indicates the
omission of any zero modes. Rather than using eq.~\scalarPI\
directly in what follows, we instead use its derivative with
respect to $m^2$, which implies:
\eq
\frac{d \Sigma}{d m^2}  = \frac{i}{2} {\Tr}^\prime \left(
\frac{1}{-\Square_g + m^2} \right) = \frac{i}{2} \int d^n x \;
\lim_{x' \to x} G(x,x'),
\label\deriveqn
\eeq
where $G(x,x')$ is the scalar Feynman propagator:
\eq
(-\Square_g + m^2)_x G(x,x^\prime) = \frac{\delta^n(x,x^\prime)}{
 \sqrt{-g}} .
\label\propdef
\eeq
To obtain the effective action we integrate eq.~\deriveqn\
with respect to $m^2$:
\eq
\Sigma (g_\mn,m^2) - \Sigma (g_\mn,m_0^2)
= \int^{m^2}_{m_0^2} dm^2 \left\{ \frac{i}{2} \int
d^n x \sqrt{-g} \; G(x,x) \right\} .
\label\massint
\eeq
The result will equal the desired effective action up to terms
independent of the mass $m^2$. The quantity $m_0$ is a reference
mass, for which we imagine the functional determinant to
have been explicitly evaluated using other means. Convenient
choices for which this is often possible are $m_0 = 0$ or $m_0 \to
 \infty$.

In this way the problem reduces to the
construction of the scalar-field Feynman propagator on
AdS spacetime, whose form in $n$ dimensions
has long been known \burgesslutken.

\section{The $n$-dimensional Effective Action}

\ref\whitandwat{Whittaker, E.T., and Watson, G.N., {\it Modern Analysis},
4th ed., Cambridge University Press, 1927, part II, 1934.}

It only remains to evaluate the previous expression using
the explicit expression for the Feynman propagator. To do
so requires a choice of vacuum state.
We work with the propagator which satisfies the
energy-conserving boundary conditions on anti-de Sitter space
\burgesslutken, which is given in terms of standard hypergeometric
functions, $F(a,b;c;x)$ \whitandwat, by:
\eq
- \, {i \over 2} \; G_\ssf(z) =
{C_{{\ssf,n}} \over 2 \; z^\beta} \; F\left(
{\beta \over 2} \; , \; {\beta+1 \over 2} \; ;\;
\beta- \, {n \over 2} + {3\over 2} \; ; \;
{z}^{-2} \right) ,
\label\exprone
\eeq
where $z=1+\lambda^2 \, \sigma(x,x')$ and $\sigma(x,x')$ is the square of the
geodesic distance between the points $x$ and $x'$, and
$\beta$ denotes the expression
\eq
\beta = \frac{n-1}{2} \pm \sqrt{
\frac{(n-1 )^2}{4}\,+\frac {m^2}{\lambda^2}  } .
\eeq
Finally, the coefficient $C_{{\ssf,n}}$ is a known constant, defined in
equation (9) of ref.~\burgesslutken:
\eq
C_{{F,n}} = \frac{ \lambda^{(n-2)}
\; \Gamma (\beta)}{2^{\beta+1} \, \pi^{n/2-1/2}
\;  \Gamma \left( \beta - \frac{1}{2} (n-3) \right)}
\label\cfexpr
\eeq

We require the coincidence limit ($\sigma \to 0$) of eq.~\exprone,
and so take $z \to 1$. Using the corresponding limit for the
hypergeometric function:
\eq
F(a,b;c;1)=
{\frac {\Gamma (c)\Gamma (c-b-a)}{\Gamma (c-b)\Gamma (c-a)}} \; ,
\eeq
and simplifying further the $\Gamma$ functions in the denominator,
the propagator's coincidence limit takes the form:
\eq
- \; {i \over 2} \; G_\ssf(1) =
{\frac {C_{{\ssf,n}} \, {2}^{\beta-n} \;
\Gamma \left(\beta- {n \over 2} +{ 3 \over 2} \right) \;
\Gamma \left(1 - {n\over 2} \right)}{\sqrt {\pi }
\; \Gamma (\beta-n+2)}} .
\eeq
When $n$ is a positive, even integer, this expression suffers from the
usual divergences that are associated with the coincidence limit
of the Feynman propagator. We regularize these by
temporarily imagining the spacetime dimension, $n$,
to be complex, with $n$ taken to the physical dimension of
spacetime only at the end of the calculation.

Combining all of these expressions,\foot\corrn{We correct here
a typo in the coincidence limit of ref.~\burgesslutken.} we find
the coincidence limit of the scalar Feynman propagator to be
\eq
-\; {i \over 2} \; G_\ssf(1) =
 \frac{ \Gamma \left( \frac n2 - \hf +
\sqrt{\frac{(n-1)^{2}}{4} +{\frac {{m}^{2}}{\lambda^2}}}\right)
\; \Gamma (1 - \, \frac n2)  \; {\lambda}^{(n-2)}
}{{2}^{n+1} \; {\pi }^{n/2} \; \Gamma \left(
-\, \frac n2 + \frac 32 + \sqrt{\frac{(n-1)^{2}}{4}
+ {\frac {{m}^{2}}{\lambda^2}}} \right) } .
\label\bigresult
\eeq

To proceed, we now integrate eq.~\bigresult\ with respect to $m^2$.
The limit $n \to D$ of eq.~\bigresult, when $D$ is an odd integer,
is well-defined and so may be taken directly, and the
result integrated with respect to $m^2$. When $D$ is even,
however, the pole from the $\Gamma$-function in the numerator
gives a divergent result, which we may isolate by
performing a Laurent series in powers of $(n-D)$. It is
generally useful to perform this expansion first, and
reserving until last the integration over $m^2$.

\section{Applications to Specific Dimensions}

We now perform the limit $n \to D$ of eq.~\bigresult\
for several choices of positive integer $D$.

\subsection{The Case $D = 2$}

Specializing to $D=2$, the Laurent expansion of the
scalar propagator
becomes (neglecting terms which are $O(n-2)$):
\eq
 \; \frac{i}{2} \; G_\ssf(1) =
\frac {1}{4 \pi \, (n-2)} \;
- \frac{1}{8 \pi} \; \left[ \ln \left( \frac{4 \pi \,
\Lambda^2}{\lambda^2} \right) - \gamma- 2 \, \Psi\left(
\hf + \hf \, \sqrt{1 + \frac{4\,m^2}{\lambda^2}}
\right) \right] ,
\label\twoD
\eeq
where $\Psi(x) := d\, \ln\Gamma(x) /dx$, $\gamma$ is
the Euler-Mascherelli constant and $\Lambda$ is
the usual arbitrary scale which enters when dimensions are
continued to complex values.

Integrating eq.~\twoD\ with respect to mass, we obtain the
effective action as the integral over an effective lagrangian
density: $\Sigma = - \int d^2x \; \sqrt{-g} \; V_\eff(\lambda^2,m^2)$, with
\eq
\eqalign{
V_\eff(\lambda^2,m^2) =& V_\eff(\lambda^2,0) -
	\left[- \, \frac {1}{4\pi \,(n-2 )} + \frac{1}{8\pi}\,
\left(- \gamma + \ln \left( \frac{4\pi \,
\Lambda^2}{\lambda^2} \right) -2 \right) \right] \; m^2 \cr
&+ \;  \frac{\lambda^2}{8 \pi}  \left[ 2 \,\ln G_1 \left(
	\frac12\,\sqrt {1+ \frac {4 m^2}{\lambda^2}} + \frac 12\right)
+ 4\,\ln  G_2\left( \frac 12 \,\sqrt {1+ \frac{4m^2}{\lambda^2}}
+ \frac12 \right) \right.  \cr
 &	\left.  + \left( 1 - \sqrt{1+ \frac{4 m^2}{\lambda^2}}
\right) \; \ln (2 \pi) \right] . \cr}
\label\exaxtmzero
\eeq

\ref\vig{M. Vigneras, Asterisque 61, (1979),  235 }

Here $G_n(x)$ denote the multiple Gamma functions, which are
defined to satisfy the following Gamma-function-like
properties:
\eq
\eqalign{
   & (1) \quad G_{n}(z+1)=G_{n-1}(z) G_{n}(z), \cr
      & (2) \quad G_{n}(1)=1, \cr
      & (3) \quad \frac{d^{n+1}}{dz^{n+1}}\log G_{n}(z+1)\geq0
         \quad  \hbox{{\rm for }} \quad z\geq0,\cr
      & (4) \quad G_{0}(z) =z \cr }
\eeq
It is a theorem \vig\ that the solutions to these conditions
are unique. Furthermore the first few functions are old friends:
$G_0(z) = z$ and $G_1(z) = \Gamma(z)$. Some useful
properties of these functions are summarized in the Appendix.

Notice, in two dimensions, that the massless reference
point is useful because the functional integral for
massless scalars is known
to give the Liouville action:
\eq
\Sigma(g_\mn,0) = -\; {1 \over 96 \pi} \int d^2 x \, \sqrt{-g} \;
R \; \left( { 1 \over \Square} \right) \; R ,
\eeq
where $\Square^{-1} \, R$ denotes the convolution of $R$
with the Feynman propagator of eq.~\propdef: $\int d^2y
\, \sqrt{-g} \; G(x,y) \, R(y)$.

Using the asymptotic expansions of the $G_n$ which are
given in the Appendix, the small curvature limit
($\lambda^2 \ll m^2$) of eq.~\exaxtmzero\ is found to be:
\eq
\eqalign{
V_\eff(\lambda^2,m^2) &\sim V_\eff(\lambda^2,0) -
\frac{m^2}{8 \pi} \left[ \frac {2}{(n-2 )}
- \ln \left( \frac{4 \pi \, \Lambda^2}{m^2} \right) +
\gamma -1\right]    \cr
& - \frac{\lambda^2}{24\,\pi} \, \left[
\ln \left( \frac {\lambda^2}{8 \pi^3 \, m^2} \right) + \frac32
-12\,\zeta^\prime(1) \right] +  \frac {\lambda^4}{120 \, \pi \,m^2}
+ O \, \left( \lambda^6 \right) , \cr}
\label\smallabsK
\eeq
where $\zeta(x)$ denotes the usual Riemann zeta function.

\subsection{The Case $D=4$}

Evaluating eq.~\bigresult\ for $n \to D=4$ dimensions permits a comparison
of this expression with previous work. 

\topic{Spinless Particles}

The expansion of eq.~\exprone\ about $n=4$ produces the following
coincidence limit:
\eq
\eqalign{
 \,  \frac{i}{2} G_\ssf & =
- \; \frac {2\,\lambda^2+m^2}{16 \pi^2 \, (n-4)}+
\frac {m^2}{32 \pi^2} \cr
&+ \left( \frac{2\,\lambda^2+m^2 }{32 \pi^2} \right) \, \left [\ln \left(
\frac{4\pi \, \Lambda^2}{\lambda^2} \right) - \gamma -2 \,\Psi \left(
\frac12 + \sqrt{\frac94 + \, \frac{m^2}{\lambda^2}}  \right)
\right ] + O(n-4) .
}
\eeq
Integrating with respect to mass then gives:
\eq
\eqalign{
V_\eff(\lambda^2,m^2) &=  V_\eff(\lambda^2,0) - \frac {\lambda^4}{64 \pi^2}
\left\{ \left( -\frac{2}{n-4} + \ln \left( \frac{4\pi \,
\Lambda^2}{\lambda^2} \right) -\gamma + \frac13 \right)
\left(b^2 - \frac94 \right) \left( b^2 + \frac74 \right)
\right.  \cr
& 	+ \left[ \left( 6 + 8 \, C_2 \right) \left(
\frac12 + b \right) - 9 +24 \, C_3 + 8\, C_2 \right]
\left( b^2 - \frac94 \right) \cr
& + \left( 24 \, C_2 + 11 +
48 \, C_3 + 48 \, C_4 \right) \left( - \; \frac32 + b \right)
\cr
& \quad \left. - 72 \,\ln G_3 \left( \frac12 +b \right)
-24 \, \ln G_2 \left( \frac12 +b \right) -48 \,
\ln G_4 \left( \frac12 +b \right) \right\} ,\cr}
\label\adsfour
\eeq
where $b^2 := \frac{9}{4} + \frac{m^2}{\lambda^2}$, and the
$C_n$ are as defined in the Appendix.

\ref\maple{Waterloo Maple 5,V4}

This expression can be compared with earlier calculations.
These have been computed in terms of the integral over
$m^2$ in ref.~\burgesslutken\ (using the same methods as
used here) and ref.~\camp\ (using $\zeta$-function methods).
The result of ref.~\camp\ is:
\eq
\eqalign{
V_\eff =  - \Scl_\eff  = & - \; \frac {\lambda^4}{64
\pi^2} \,\left[ \left( {b}^{4} -\frac{1}{2} \,{b}^{2}-
\frac{17}{240} \right) \ln\left( \frac
{\nu^2}{\lambda^2} \right) + {b}^{4} +
\frac{1}{6} \,{b}^{2} +8\,c \right] \cr
& \quad + \;
\frac{\lambda^4}{16 \pi^2} \,\int_{1/2}^{1/2+b} \! x\left(
x-1\right)\left(2\,x-1\right)\Psi(x) \; {dx} \cr}
\label\campzero
\eeq
where $\nu$ is the arbitrary scale which arises in $\zeta$-function
regularization, and the constant $c$ is given
by \foot\typoc{We correct here a typo in ref.~\camp, where
the value for the constant $\ss c$ is incorrect by $\ss - 137/360$ }
\maple:
\eq
\eqalign{
c &= \int_0^\infty  \frac{2 u\,
\left( {u}^{2}+1/4
\right )\ln u }{e^{2\pi\, u}+1} \; d u \cr
&=
- \; { \frac{\ln 2}{160}} - \frac {17}{960} \, \ln \pi +
\frac{137}{5760} - \frac{17}{960} \, \gamma +
\frac{21}{32} \, \frac{\zeta^\prime (4)}{\pi^4}
+\frac{1}{16} \, \frac{\zeta^\prime (2)}{\pi^2} \cr
&= -0.01744158583... \cr}
\eeq
If we evaluate the integrals in eq.~\campzero\ in terms
of the multiple Gamma functions, and subtract
the result for $m=0$ limit, we find agreement with
eq.~\adsfour, provided the arbitrary scales $\nu$ and $\Lambda$
are related in the following way:
\eq
\Lambda = \nu \, \exp \left[ \frac{ (12 b^2 + 21)
(\gamma - \ln (4 \pi))+
56}{6(4 b^2 + 7)}
\right]
\eeq

\topic{Higher Spins for $D=4$ Anti-de Sitter Space}

Some results are also available in four dimensions for higher-spin
particles. It is often possible to express the one-loop functional
determinants for higher-spin fields in the form
\eq
\det \Bigl( - \Square_s + X \Bigr)
\label\hispindet
\eeq
where $\Square_s$ is the Laplacian operator acting on various
constrained tensor and/or spinor fields. (For instance, for spin-1
particles the relevant field is a divergenceless vector
field.) The functional determinants for these fields have been
evaluated for dS spacetimes in ref.~\allen, and for AdS
spacetimes in ref.~\camphig, using $\zeta$-function
regularization. Following these references, we label these
fields by the corresponding spin, $s$, where $s$ is an integer
for tensors and a half-odd integer for spinors.
For tensor fields ($s = \hbox{integer}$) on AdS with $D=4$
ref.~\camphig\ gives the following result (with the overall
sign chosen for bose statistics):
\eq
\eqalign{
V_\eff^s &=  - \; g(s) \; {\frac {\lambda^4}{64 \pi^2}}
\left\{ \left[ b^4 - \left( s + \frac{1}{2} \right)^{2}
\left(2 \,{b}^{2} + \frac{1}{6} \right) - \frac{7}{240}
\right] \ln \left( \frac{\nu^2}{\lambda^2} \right) +
b^4 + \frac{1}{6} \, b^2 + 8 \, c_+ \right\}
\cr
& \qquad\qquad\qquad -g(s) \;
\frac{\lambda^4}{8 \pi^2} \int _{0}^{b} \! \left[
\left( s + \frac{1}{2} \right)^{2} - {t}^{2} \right]
\; \Psi\left( t+ \frac{1}{2} \right) \, t \; dt ,\cr}
\label\bosons
\eeq
with $g(s) = 2s + 1$. The quantity $b$ is given in refs.~\camphig\
and \allen, and depends on both $m^2/\lambda^2$ and $s$. For the special
case $s = 0$ we have $b^2 = \frac{9}{4} + \frac{m^2}{\lambda^2}$,
while for $s=1$, $b^2 = \frac{1}{4} + \frac{m^2}{\lambda^2}$.
The constant $c_+$ is given by \maple,
\eq
\eqalign{
c_+ &= \int_{0}^{\infty } \;\; \frac{ 2  u
\, \left[  u^2 + \left(s + \frac{1}{2} \right)^2
\right] \ln  u }{e^{2\pi \, u} +1 }\; d
 u \cr
=&  \frac{s(s+1)}{24} \left(-\,\ln \pi +1 - \,\gamma
+ \frac{ 6\, \zeta^\prime(2)}{\pi^2} \right)
- \frac{\ln 2}{160}  -  \frac{17 \, \ln\pi }{960} \cr
& \quad + \frac{137}{5760} - \frac{17 \, \gamma}{960}
+  \frac{21 \, \zeta^\prime(4)}{32 \pi^4} +
 \frac{\zeta^\prime(2)}{16 \pi^2} . \cr}
\eeq
Evaluating the integrals in eq.~\bosons\  we find
the effective Lagrangian produced by (constrained)
tensor fields on AdS expressed in terms of the multiple
Gamma functions:
\eq
\eqalign{
V_\eff &= g(  s) \; \frac{ \lambda^4}{64 \pi^2}  \left\{
\left[ \ln \left( \frac{\lambda^2}{\nu^2} \right) - \frac13
\right] b^4 - \left( 8\,C_2 + 6\right) b^3 \right. \cr
& \qquad\qquad + \left[ -2 s(s+1) \left( 1 + \ln \left(
\frac{\lambda^2}{\nu^2} \right) \right)  -24 \,C_3
+ \frac{3}{2} -12\,C_2 - \frac{1}{2} \,\ln \left(
\frac{\lambda^2}{\nu^{2} } \right) \right] b^2 \cr
& \qquad\qquad + \left[ 2s (s+1) \left( 4\, C_2 +1 \right)
 -48\,C_3 + \frac{5}{2} - 48\, C_4 - 6\,C_2 \right] \; b \cr
&  \qquad\qquad\;+ \left[ - \frac{ 1}{6}
\,\ln \left( \frac{\lambda^2}{\nu^2} \right) +4 \,
\ln G_1 \left( \frac{1}{2} \right) + 8\, \ln G_2
\left( \frac{1}{2} \right) -4 \, \ln G_1 \left(
\frac{1}{2}+b \right) \right. \cr
&  \qquad\qquad\; \left. -8\, \ln G_2 \left( \frac{1}{2}
+b \right) \right] s (s+1 )
+24 \, \ln G_2 \left( \frac{1}{2} +b \right) +
72\, \ln G_3 \left( \frac{1}{2} +b \right) -8
\, c_+ \cr
&  \qquad\qquad\;  -24 \, \ln G_2 \left( \frac{1}{2} \right)
- \frac{17}{240} \, \ln \left( \frac{\lambda^2}{\nu^2} \right)
-48 \, \ln G_4 \left( \frac{1}{2} \right)
 -72 \, \ln G_3 \left( \frac{1}{2} \right) \cr
& \qquad\qquad\; \left. +48\,\ln G_4 \left(
\frac{1}{2}+b \right) \right\} \qquad\qquad
\qquad\qquad \hbox{for} \; s = \hbox{integer}. \cr}
\label\bosontwo
\eeq

A similar result may be derived for (constrained) spinor fields.
Ref.~\camphig\ gives the following expression (assuming fermi
statistics):
\eq
\eqalign{
V_\eff^s &= g(  s) \; \frac{\lambda^4}{64 \pi^2} \, \left\{
\left[ {b}^{4} - \left( s+ \frac{1}{2} \right)^2
\left( 2\, {b}^{2} - \frac{1}{3} \right) + \frac{1}{30}
\right] \ln \left( \frac{\nu^2}{\lambda^2} \right)  \right. \cr
& \qquad \qquad \left. + {b}^{4} - \frac{4 \, b^3}{3}
- \frac{b^2}{3}   + 4\,\left( s+\frac{1}{2} \right )^2
b - 8\, c_- \right\} \cr
& \qquad\qquad + g(  s) \; \frac{\lambda^4}{8 \pi^2}   \int_{0}^{b
} \! \left[ \left( s+ \frac{1}{2} \right)^{2} - {t}^{2}
\right] \; \Psi(t) \, t\;  dt , \cr}
\label\fermion
\eeq
where $b$ is again spin dependent, equal to $b^2 =
\frac{m^2}{\lambda^2}$ for $s=\frac{1}{2}$. The constant $c_-$
is \maple:
\eq
\eqalign{
c_- &= \int_{0}^{\infty } \;\; \frac{ 2  u
\left[  u^2 + \left( s+ \frac{1}{2} \right)^{2}
\right] \ln  u }{ e^{2\pi \, u} -1} \; d u \cr
& = - \frac{7 \, \ln 2}{240} - \frac{7 \, \ln \pi}{240}
+ \frac{13}{360} - \frac{7 \, \gamma}{240} +
\frac {3 \, \zeta^\prime (4)}{4\pi^4} \cr
& \qquad\qquad + \frac{s(s+1)}{12} \, \left[ - \, \ln (2 \pi)
+1 - \, \gamma + \frac{6\, \zeta^\prime (2)}{\pi^2}
\right] + \frac{1}{8} \, \frac{\zeta^\prime (2)}{\pi^2}.\cr}
\eeq
Combining expressions we find the following form for the
spinor effective Lagrangian on AdS:
\eq
\eqalign{
V_\eff^s &= g(  s) \; \frac { \lambda^4}{64 \pi^2} \, \left\{
\left[ -\ln \left( \frac{\lambda^2}{\nu^2} \right) -
\frac{13}{3} \right] \; {b}^{4} + \left( 64\, C_2
+ \frac{124}{3} \right) \; {b}^{3} \right.  \cr
& \qquad + \left[ \left( 16+ 2\, \ln \left( \frac{\lambda^2}{\nu^2}
\right) \right) s(s+1)  + \frac{1}{2} \, \ln \left(
\frac{\lambda^2}{\nu^2} \right) - \frac{101}{3} + 96\,C_2
+ 192 \,C_3 \right] \; {b}^{2} \cr
& \qquad \qquad + \left[ \phantom{\frac12}
(-64\, C_2 -28 ) s(s+1) -39
 +384\, C_{{3}} +384\,C_{{4}} + 48 \,C_{{2}} \right] \; b \cr
& \qquad \qquad +\left[ 64\, \ln G_{{2}} (b) - \frac{1}{3} \, \ln
\left( \frac{\lambda^2}{\nu^2} \right) + 64 \, \ln G_{{1}} (b)
\right] s(s+1)  \cr
& \qquad \qquad -768 \,\ln G_{{3}}(b)- {\frac {7}{60}}
\,\ln \left( \frac{\lambda^2}{\nu^2} \right) - 8 \,c_-
-432\, \ln G_{{2}}(b) \cr
&\qquad\qquad \left. -48\, \phantom{\frac12}
\ln G_{{1}}(b)- 384\,\ln G_{{4}}(b) \right\} \qquad
\hbox{for} \qquad s = \hbox{half-integer}. \cr}
\label\fermionaction
\eeq
The following technical point bears notice. When evaluated
for massless, spin $\frac{1}{2}$ fermions
($b=0$), eq.~\fermionaction\ superficially
appears to be ill-defined, due to the appearance of the divergent
quantities $\ln G_2(0)$, $ \ln G_3(0)$ and $\ln G_4(0)$.
It happens that these divergences cancel in
eq.~\fermionaction, leaving a well-defined massless limit.

\endtopic

\subsection{Scalar Fields in Odd Dimensions}

We now turn to the effective action for massive scalar
fields in odd-dimensional anti-de Sitter spacetimes.
As is usually the case for dimensionally-regularized
one-loop quantities, the resulting expressions
are easier to evaluate due to the absence in odd
dimensions of logarithmic divergences at one loop.

We simply quote here the final results for the effective
lagrangian for the lowest odd dimensions.

\topic{$D = 3$}

For 3-dimensional AdS spacetimes the massive scalar effective
lagrangian density becomes:
\eq
V_\eff(K,m) - V_\eff(K,0) = - \; \frac{\lambda^{3}}{12 \pi} \;
\left[ \left( \frac {\lambda^2+{m}^{2}}{\lambda^2} \right)^{3/2}  -
 1 \right] .
\eeq

\topic{$D=5$}

The corresponding result for 5-dimensional AdS spacetimes is:
\eq
V_\eff(K,m) - V_\eff(K,0) =   \frac {\lambda^{5}}{360 \,\pi^2}
\,\left[ \left( \frac {4\,\lambda^2+m^2}{\lambda^2} \right)^{3/2}
\left( \frac{ 7\,\lambda^2+3\,m^2}{\lambda^2} \right) - 56 \right] .
\eeq

\topic{$D=7$}

For $D=7$ we have:
\eq
\eqalign{
&V_\eff(K,m) - V_\eff(K,0) =  \cr
&\qquad\qquad\qquad -\;  \frac {\lambda^{7}}{5,040 \, \pi^3}
 \,\left[ \left( \frac{9\,\lambda^2+m^2}{\lambda^2} \right)^{3/2}
\left( \frac{ 82\,{\lambda^2}^{2} +33\,\lambda^2{m}^{2} +3\,{m}^{4}}
{\lambda^4} \right) - 2,214 \right] . \cr}
\eeq

\topic{$D=9$}

For $D=9$:
\eq
\eqalign{
&V_\eff(K,m) - V_\eff(K,0) =  \frac{\lambda^{9}}{151,200 \, \pi^4}\,
\left[ \left( \frac{16\,\lambda^2+m^2}{\lambda^2}
\right )^{3/2} \right. \cr
& \qquad \qquad \left. \times  \left( \frac{3,956\,
\lambda^6+1401\,\lambda^4 m^2+150\,\lambda^2 m^4 +
5\,m^6}{\lambda^6} \right)
 - 253,184 \right]. \cr}
\eeq

\topic{$D=11$}

Finally, the 11-dimensional expression is:
\eq
\eqalign{
&V_\eff(K,m) - V_\eff(K,0) =
 - \;  \frac {\lambda^{11}}{1,995,840 \, \pi^5} \,
\left[ \left( \frac{25\,\lambda^2+{m}^{2}}{\lambda^2}
\right)^{3/2} \right. \cr
& \; \left. \times
\left( \frac{128,536\,{\lambda}^{8}+40,188\,{\lambda}^{6}{m}^{2}
+4,287\,{\lambda}^{4}{m}^{4}
+190\,\lambda^2{m}^{6}+3\,{m}^{8} }{\lambda^8} \right)
- 16,067,000 \right]. \cr}
\eeq

\endtopic

\vfil\eject
\section{Duality}

Recently, Cruz \claim\ has proposed the classical equivalence of
two types of free scalar fields in two-dimensional AdS spacetime. 
The proposed equivalence relates a massless, minimally-coupled
scalar with a massive scalar having mass $m^2 = R = 2 \lambda^2 $.
He argues for this equivalence by constructing a
time-dependent canonical transformation which maps
one system into the other.

\ref\dhokerjackiw{E. D'Hoker, R. Jackiw, {\it Classical and quantal
Liouville field theory}, \prd{26}{82}{3517}}
\ref\dhoker{E.D'Hoker, {\it Equivalence of Liouville Theory
and 2-D Quantum
Gravity}, \mpla{6}{91}{745}   }

In this section we wish to argue against the existence of
this equivalence at the quantum level.  Of course, 
the absence of a quantum symmetry need not preclude
the existence of a classical symmetry.  The failure
of a canonical transformation to survive promotion to the quantum theory is
similar to what happens for the Liouville action, which is
canonically equivalent to a free field theory --- and
so is integrable \dhokerjackiw\ ---  but is nonetheless quantum
mechanically distinct from it (see, ref.~\dhoker,
and references therein).

In defense of our point of view we use the calculations
of the previous section to see if duality is maintained at the quantum
level. One would expect equivalence to imply the equality of
the effective actions $\Sigma$ computed for the two types of scalars.
This amounts to the vanishing of expression \exaxtmzero, which gives
the difference between the massive and massless effective potentials.
Since the arguments of ref.~\claim\ apply for any $ \lambda^2  > 0$,
eq.~\exaxtmzero\ should vanish for  all such $\lambda^2$. We find:
\eq
\eqalign{
V_\eff(\lambda^2,m^2) - V_\eff(\lambda^2,0) &= -
	\left[C + \frac{1}{8\pi}\,\ln \left( \frac{
\Lambda^2}{ \lambda^2 } \right) \right] \; m^2
 +  \frac{ \lambda^2 }{8 \pi}  \; \left[
\left( 1 - \sqrt{1+ \frac{4 m^2}{ \lambda^2 }} \right)
\; \ln (2 \pi)  \right. \cr
& \qquad \left. +  2 \,\ln G_1 \left(
	\frac12\,\sqrt {1+ \frac {4 m^2}{ \lambda^2 }} + \frac 12\right)
+ 4\,\ln  G_2\left( \frac 12 \,\sqrt {1+ \frac{4m^2}{ \lambda^2 }}
+ \frac12 \right)  \right] ,\cr}
\label\deltatildeV
\eeq
where $C$ is the contribution of any counterterms.
Besides cancelling the divergence of eq.~\exaxtmzero\ as
$n\to 2$, these depend on $\L$ in just such a way as to ensure
the $\L$-independence of $V_\eff$.
Evaluating this expression for $m^2 = 2 \lambda^2 $ we find
\eq
V_{eff}(\lambda^2,m^2 = 2\lambda^2) - V_{eff}(\lambda^2,0)
 = - \left[ 2 \, C + \frac{1}{4 \pi}
\ln \left( \frac{2 \pi \,\Lambda^2}{ \lambda^2 } \right)
\right]  \;  \lambda^2   ,
\label\dualspcase
\eeq
where we have used $G_1(2) = G_2(2) = 1$.

Clearly, so long as $C$
may depend arbitrarily on $\lambda^2$ and $m^2$, we are always
free to choose $C$ to ensure the vanishing of
eq.~\dualspcase. $C$ may certainly depend
on $\lambda^2$, since the counterterms can involve powers
of the curvature, $R$.

(The reader might wonder
why we entertain here the possibility of curvature-dependent
counterterms when, for the noninteracting scalar
on a fixed gravitational background under consideration,
we have seen that no $\lambda^2$ dependence is required to cancel
divergences in two dimensions. We do so because
more complicated counterterms {\it are} required
once interactions are included, and if the gravitational
field is also treated as a quantum field. Moreover,
we must consider the possibility that
duality at the quantum level may require special
choices for finite counterterms, even if these are not
required to cancel divergences.)

We now come to the main point. There are now two ways
to proceed, depending on how much
$\lambda^2$ dependence we are prepared to entertain.

\topic{Option 1: Arbitrary $\lambda^2$ Dependence}

One way to proceed is to damn the torpedoes
and to permit $C$ to depend arbitrarily on $\lambda^2$.
This might be reasonable if we regarded the metric
strictly as a background field, and permitted
the addition to the classical action of an
arbitrary metric-dependent functional which
is independent of our scalar field, $\phi$.
In this case, in the interest of enforcing
a quantum duality, we choose $C$ to ensure the
vanishing of eq.~\dualspcase\ for all $\lambda^2$. With
this choice, eq.~\deltatildeV\ becomes:
\eq
\eqalign{
V_\eff(\lambda^2,m^2) - V_\eff(\lambda^2,0)  =& \frac{m^2}{8 \pi}
\; \ln (2 \pi)
+ \;  \frac{ \lambda^2 }{8 \pi}  \left[
\left( 1 - \sqrt{1+ \frac{4 m^2}{ \lambda^2 }}
\right) \; \ln (2 \pi) \right. \cr
&\; + \left.  2 \,\ln G_1 \left(
	\frac12\,\sqrt {1+ \frac {4 m^2}{ \lambda^2 }} + \frac 12\right)
+ 4\,\ln  G_2\left( \frac 12 \,\sqrt {1+ \frac{4m^2}{ \lambda^2 }}
+ \frac12 \right)  \right]  .\cr}
\label\dualrenV
\eeq

Eq.~\dualrenV\ is plotted in Figure 1,  using the
variables $y = [V_\eff(\lambda^2,m^2) - V_\eff(\lambda^2,0)]/ \lambda^2$
\vs\ $x = m^2/2 \lambda^2 $.
The following points emerge from an inspection of this plot.

\item{1.} By construction $y(0) = y(1) = 0$ indicating the
equivalence of $V_\eff$ when evaluated at $m^2 = 0$ and
$m^2 = 2 \lambda^2 $.
But the construction just given shows that there is
nothing special about
the choice $m^2 = 2 \lambda^2 $, since we could have
equally well renormalized to
ensure $y=0$ for some other value of $m^2$.

\item{2.}
Because $y(x)$ is not monotonically increasing or decreasing,
there are many pairs $\{x_1,x_2\}$ which satisfy $y(x_1) = y(x_2)$, and
so many pairs $\{m^2_1,m^2_2\}$ for which $V_\eff$ takes the same value.
What is less obvious from the plot, but nevertheless true,
is that the slope, $\partial V_\eff/\partial
\lambda^2$, is not
the same for both members of these pairs. Since these slopes are
related to the expectation $\Avg{ {T^\mu}_\mu}$ for the scalar
field stress-energy tensor, this quantity must differ for $m_1$ and $m_2$
even though $V_\eff$ takes the same value for these two masses.

We conclude that duality is not a property of
the quantum theory.

\topic{Option 2: Polynomial $\lambda^2$ Dependence}

A more reasonable requirement on $C$, in our opinion,
is to require it to be at most a polynomial in
$\lambda^2$ (to any fixed order in perturbation theory).
Physically, counterterms arise once higher-energy
physics is integrated out, and so they should be
interpreted in an effective-lagrangian sense. That is,
they should be treated as perturbations in a low-energy
derivative expansion. If so, to any fixed order in
this expansion, they
must be generally-covariant powers of the fields
$\phi$ and $g_\mn$ and their derivatives,
restricting $C$ to be a polynomial in $\lambda^2$.

If so, it is no longer possible to choose $C$ to
ensure the vanishing of eq.~\dualspcase\ because
cancellation would require $C$ to depend logarithmically
on $\lambda^2$. Once again we are led to conclude that duality
does not survive quantization.

\endtopic

\section{Acknowledgements}

We would like to acknowledge
the Instituto de F\'isica, Universidad Nacional
Aut\'onoma de M\'exico, for their kind
hospitality during part of this work. Our research is
funded in part by the Natural Sciences and
Engineering Research Council of Canada (NSERC),
as well as the McGill-UNAM exchange agreement.

\appendix{A}{$G_n$: the Multiple Gamma function}

In this appendix we state some principal formulae pertaining to the
multiple gamma function. We also derive an integral representation for
these functions, and use it to obtain closed forms for the integral
moments of the $\Psi(x) = d \ln \Gamma(x) / dx$ function.

\ref\uenonishi{K. Ueno, M. Nishizawa,  q-alg/9605002}

\topic{Defining Properties }

\ref\BA{E. Barnes, Quaterly Journal of Mathematics, Vol. 31,(1899) ,264  }

In 1900, Barnes \BA\ introduced a generalization of the $\Gamma$ function,
denoted $G(x)$, which satisfies:
\eq
G(z+1)=\left (2\,\pi \right )^{1/2\,z}{e^{-1/2\,z\left (z+1\right )-1/
2\,\gamma\,{z}^{2}}}\prod _{n=1}^{\infty }\left (1+{\frac {z}{n}}
\right )^{n}{e^{-z-1/2\,{\frac {{z}^{2}}{n}}}}
\label\barnesdef
\eeq
and which satisfies the properties $G(z+1)= \Gamma (z)G(z)$
and $G(1)= 1$.

This was further generalized by
Vign\'{e}ras \vig\ in 1979, who introduced a hierarchy of
Multiple Gamma functions, $\{ G_n \} $, for $n = 0,1,2,\dots$.
These functions may be defined using the following theorem.

{\bf Theorem \vig:}
There exists a unique hierarchy of functions which satisfy
\eq
\eqalign{
      & (1) \quad G_{n}(z+1)=G_{n-1}(z) G_{n}(z), \cr
      & (2) \quad G_{n}(1)=1, \cr
      & (3) \quad \frac{d^{n+1}}{dz^{n+1}}\log G_{n}(z+1)\geq0
         \quad  { for } \quad z\geq0,\cr
      & (4) \quad G_{0}(z) =z \cr
\label\vignerasthrm }
\eeq

The first three elements of this sequence of functions are
then $G_0(z) = z$, $G_1(z) = \Gamma(z)$ and $G_2(z) = G(z)$,
with $G(z)$ as defined in eq.~\barnesdef.

\topic{`Stirling' Formulae}

Vigneras \vig\ derived a Weistrass product representation
for the multiple gammas.
Another infinite product representation is derived by
Ueno and Nishizawa in
\uenonishi. They also derive asymptotic expansions
for general $G_n$,
which are the analogues of the Sterling formula for
the $\Gamma$ function. We
quote \uenonishi\ for some of these results for
low values of $n$.

In the case $n=1$, we have the usual Stirling formula for large $z$:
\eq
\eqalign{
    \log G_{1}(z+1)&=\log \Gamma(z+1)\cr
     & \sim
      \left(z+\frac{1}{2}\right)\log(z+1) - (z+1) - \zeta'(0)
      + \sum_{r=1}^{\infty} \frac{B_{2r}}{[2r]_{2}} \;
    \left(\frac{1}{z+1}\right)^{2r-1} ,
}
\eeq
where $[2r]_{n}$ stands for $\Gamma(2r + 1) / \Gamma (2r - n + 1)$.
The generalization to $n=2$, first derived by Barnes \BA, is:
\eq
\eqalign{
      \log G_{2}(z+1)
     &  \sim
       \left(\frac{z^{2}}{2}-\frac{1}{12}\right) \log(z+1)
       -\frac{3}{4}z^{2}-\frac{z}{2}+\frac{1}{4}
      - z \zeta'(0)+ \zeta'(-1) \cr
     & \quad -\; \frac{1}{12 (z+1)}
       + \sum_{r=2}^{\infty} \frac{B_{2r}}{[2r]_{3}}
       \; \frac{(z-2r+1)}{(z+1)^{2r-1}}
}
\eeq
For $n=3$ and $n=4$, the asymptotic expansions are as follows:
\eq
\eqalign{
   \log G_{3}(z+1)
   &  \sim
     \left(\frac{z^{3}}{6}-\frac{z^{2}}{4}+\frac{1}{24}\right)\log(z+1)
     - \frac{11\, z^{3}}{36}+\frac{5\, z^{2}}{24}
     + \frac{z}{3} - \frac{13}{72}\cr
   &  \quad - \frac{z(z-1)}{2} \; \zeta'(0)
     + \frac{2z-1}{2} \; \zeta'(-1) -\frac{1}{2}\;\zeta'(-2)\cr
   & \quad  + \frac{1}{12 (z+1)}
     +  \sum_{r=2}^{\infty}
      \frac{B_{2r}}{[2r]_{4}} \;
     \frac{\left[ z^{2} - (6r-11)z+(4r^2-16r+16)\right]}{(z+1)^{2r-1}}
      \cr
  \log G_{4}(z+1)
  & \sim
    \left(\frac{z^4}{24}-\frac{z^3}{6}+\frac{z^2}{6}
    -\frac{19}{720}\right)\log(z+1) - \frac{4\, z^4}{72}
    +\frac{2\, z^3}{9}
    + \frac{z^2}{8}-\frac{11\, z}{36} +\frac{31}{144}\cr
  & \quad - \frac{z^3 - 3z^2 +2z}{6} \;\zeta'(0)
    + \frac{3z^2-6z+2}{6} \; \zeta'(-1)
    -\frac{z-1}{2} \; \zeta'(-2)+\frac{1}{6} \; \zeta'(-3)\cr
  & \quad - \; \frac{1}{12 (z+1)}+
   \frac{1}{720\, (z+1)^3}\left(
    6z^2+\frac{13\, z}{2} +\frac{5}{2}\right)
  + \sum_{r=3}^{\infty}\frac{B_{2r}}{[2r]_{5}} \;
    \frac{N(z)}{(z+1)^{2r-1}} , \cr
\hbox{where} \qquad N(z) &:= \left[
    z^3-(12r-27)z^2+(20r^2-94r+111)z
    -(8r^3-56r^2+134r-109)
    \right].\cr }
\eeq

\topic{Integral Representations}

Next, we prove the following line integral representation
of the logarithm
of the multiple Gammas.

{\bf Theorem:}
\eq
\ln G_n(z+1) =  \int_0^\infty dt \,
\frac{e^{-t}}{t} (-1)^n \, \left[ \frac{1 - e^{-zt}}{(
1 - e^{-t} )^{n} } +
\sum_{m=1}^n \frac{(-1)^m }{ (1 - e^{-t} )^{n-m} } { z \choose m}
 \right]
\label\linrep
\eeq
{\bf Proof:} We show explicitly that the defining conditions in
\vignerasthrm\ are satisfied. The proof follows by induction on $n$ and from
the uniqueness of the hirarchy of $\{G_n\}$ \vignerasthrm.

i) $\ln G_n(z+2) = \ln G_{n-1}(z+1) + \ln G_{n} (z+1) $ follows from the
binomial relation:
\eq
{ z+1 \choose m} = { z \choose m-1}  +  { z \choose m}
\eeq
The integrand splits up as follows:
\eq
\eqalign{
(-1)^n \, & \left(
	\frac{1 - e^{-zt}e^{-t}}{(1 - e^{-t} )^{n} } +
		 \sum_{m=1}^n  \frac{(-1)^m }{ (1 - e^{-t} )^{n-m}} {z+1 \choose m}
\right)   \;   \cr
& \qquad \qquad\qquad= \;  (-1)^n \, \left( \frac{1 - e^{-zt}}{(1 - e^{-t}
)^{n} }+ \sum_{m=1}^n \frac{(-1)^m }{ (1 - e^{-t} )^{n-m}} {z \choose m}
\right) \cr
& \qquad\qquad \qquad\quad + (-1)^{n-1} \, \left( \frac{1 - e^{-zt}}{(1 -
e^{-t} )^{n-1} } + \sum_{m=1}^{n-1} \frac{(-1)^m }{ (1 - e^{-t} )^{n-m-1}}
{z \choose m} \right)}
\eeq
where the index on the second sum has been shifted to bring it to the
standard form.

ii) $\ln G_n(1) = 0$ follows from the vanishing integrand in the limit $z
\rightarrow 0 $;

iii) $(d/dz)^{n+1} \ln G_n(z+1) \geq 0$ follows from the absolute
positivity of the integrand:
\eq
\int_0^\infty e^{-t} \; \left[\frac{- (-t)^n+1}{(1 - e^{-t} )^{n} } \right]
\; \frac{dt}{t} \geq 0
\eeq

iv) Setting $n \rightarrow 0$ reduces to an integral representation of
$\ln(z+1)$ and $n \rightarrow 1$ to a standard representation of the
logarithm of the $\Gamma$ function, thereby completing the proof by
induction on $n$.

{\bf Corollary:} Using the integral representation of $G_n$ we derive the
following tower of relations among the logarithmic derivatives $\psi_n(z+1)
:= d \ln G_n(z+1)/dz$:
\eq
\eqalign{
\psi_2(z+1) - z \;\psi_1 (z+1) &= C_2- \frac{z}{2}  \cr
\psi_3(z+1) - z \;\psi_2(z+1) + \frac{z(z+1)}{2!} \;\psi_1(z+1)&= C_3 +
\frac{3\, z}{4} + \frac{z^2}{4} \cr}
\eeq
and
\eq
\eqalign{
&\psi_4(z+1) - z \;\psi_3(z+1) + \frac{z(z+1)}{2!} \;\psi_2(z+1)
-\frac{z(z+1)(z+2)}{3!} \;\psi_1(z+1)  \cr
& \qquad\qquad\qquad\qquad\qquad\qquad\qquad= C_4
-\frac{11 \, z}{18}-\frac{z^2}{3} -\frac{z^3}{18} .\cr }
\label\psirelations
\eeq
where $C_2 := - \; \zeta^\prime(0) - \frac{1}{2} = \frac{1}{2}
 [\ln(2\pi)-
1] = 0.4189385...$, $C_3 := -.3332237448...$, $C_4 := .2786248832...$,
 \etc.

{\bf Corollary:} Substituting lower order relations in the higher order
ones, and integrating with respect to $z$, we find
\eq \label\intzero
\eqalign{
&\int^{a}\; z\; \psi_{{1}} (z+1) \;{dz} =\ln
G_{{2}}(a+1)- a \, C_{{2}} +\frac{{a}^{2}}{4}  \cr
&\int^{a}\; \frac{1}{2!} \, z (z-1) \; \psi_{{1}}(z+1)
\; {dz}=\ln G_{{3}}(a+1)+ \frac{{a}^{3}}{12} -\left (\frac{C_{{2}}}{2}
+\frac{3}{8}\right){a}^{2}- a \, C_{{3}}\cr
&\int^{a}\; \frac{1}{3!} \,z(z-1)(z-2) \; \psi_{{1}}(z+1) \; {dz
}=\ln G_{{4}}(a+1)+ {\frac {{a}^{4}}{72}} - \left
(\frac{C_{{2}}}{6}+\frac{2}{9} \right){a}^{3} \cr
& \quad  \quad  \quad  \quad  \quad  \quad  \quad  \quad  \quad  \quad
\quad  \quad  \quad  \quad  \quad  -\left ( \frac{C_{{3}}}{2}-{\frac
{11}{36}}-\frac{C_{{2}}}{4}
\right ){a}^{2}- a \, C_{{4}}
}
\eeq

The integrals \intzero\ may be rewritten as follows:
\eq
\int^a \; {z}^{n}\psi(z+1) \; {dz}=   \left\{
	\matrix{ 	n=0: &   	\ln  G_{{1}}(a+1)  \hfill	\cr
			n=1: & 	\ln  G_{{2}}(a+1) -				aC_{{2}}+\frac{1}{4}\,{a}^{2}\hfill		\cr
			n=2: & 	\frac{1}{6}\,{a}^{3}+\left (-\frac{1}{2}-C_{{2}}\right
){a}^{2}+\left (-C_{{2}}-2\,C_{
				{3}}\right )a+
				2\,\ln G_{{3}}(a+1) +	\hfill	\cr
			& \quad \ln  G_{{2}}(a+1) 						\hfill	\cr
			n=3: &	\frac{1}{12}\,{a}^{4}+\left (-C_{{2}}-\frac{5}{6}\right
){a}^{3}+\left (-\frac{1}{6}-\frac{3}{2}\,C_{{
2}}-3\,C_{{3}}\right ){a}^{2}	+	\hfill	\cr
& \quad \left (-6\,C_{{3}}-C_{{2}}-6\,C_{{4}}
\right )a+	6\,\ln G_{{4}}(a+1)+\hfill		\cr
		& \quad 6\,\ln G_{{3}}(a+1)+\ln G_{{2}}(a+1
)
	\hfill	\cr
		} \right.
\label\intprime
\eeq
%


\vfill\eject
\listrefs



\vfill\eject
\centerline{\epsfxsize=11.5cm\epsfbox[45 430 550 750]{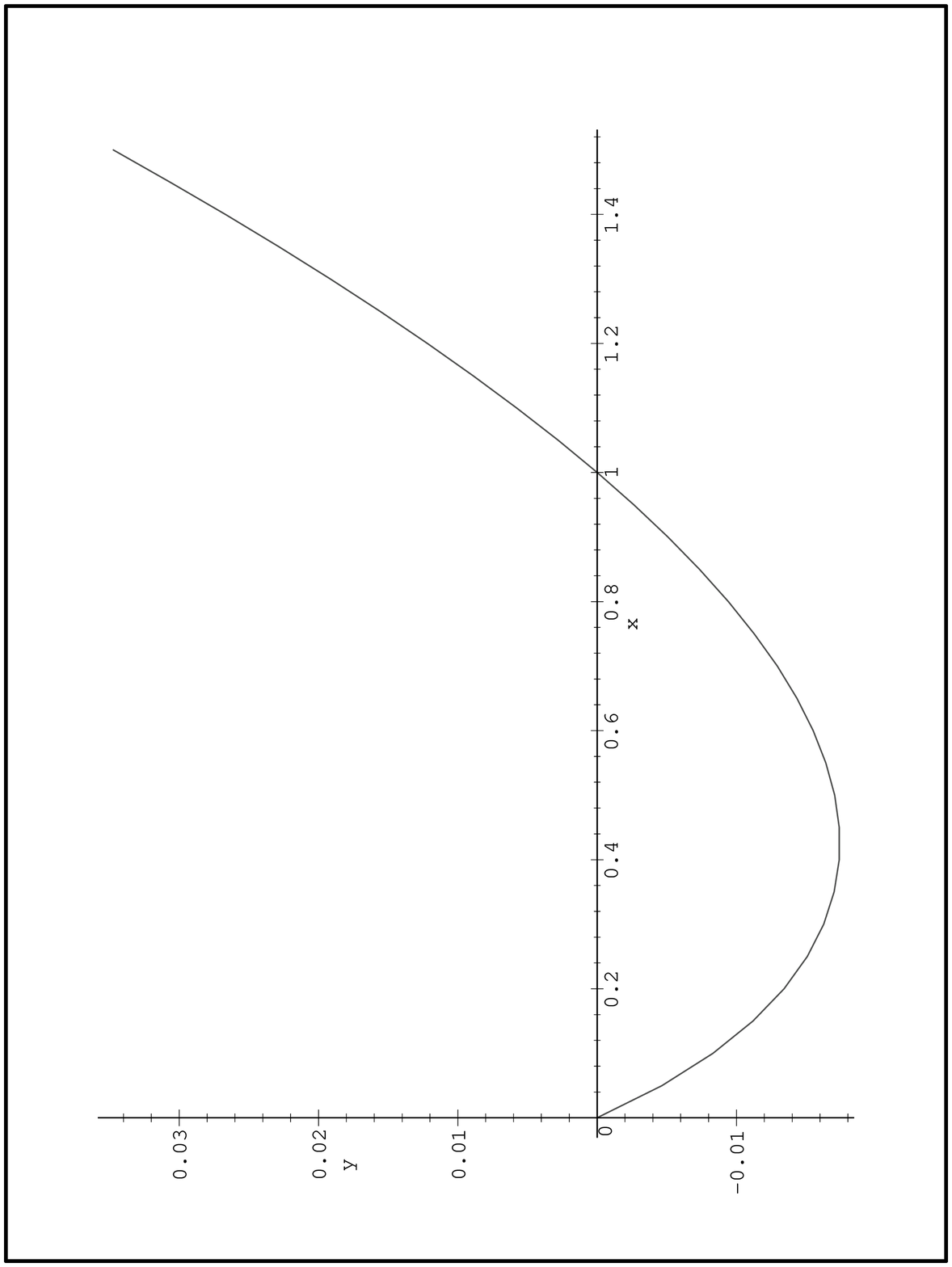}}

\vskip 5truein
\centerline{{\bf Figure 1:} A plot of $y = [V_\eff(\lambda^2,m^2) -
V_\eff(\lambda^2,0)]/ \lambda^2$
\vs\ $x = m^2/2 \lambda^2 $.}

\bye